\documentclass[twocolumn, trackchanges, 12pt fonts]{aastex631}
\usepackage{graphicx,latexsym,amssymb, epsfig}
\usepackage{multirow,amsmath,array,booktabs}
\usepackage{subfigure}
\usepackage{tabularx}
\usepackage{color}
\usepackage{bm}
\usepackage[figuresright]{rotating}
\usepackage[section]{placeins}
\usepackage{slashed}
\usepackage{graphicx}
\usepackage{tikz}

\newcommand{\nc}{\newcommand}       
\nc{\vc}[1] {\mbox{\boldmath $#1$}} 
\nc{\del}       {\partial}              
\nc{\bra}       {\langle}               
\nc{\ket}       {\rangle}               
\nc{\bras}[1]   {\langle #1|}           
\nc{\kets}[1]   {|#1\rangle}            
\nc{\mapleft}[1]{           
	\smash{\mathop{\,          %
			\hbox to 1.5cm{\rightarrowfill}\, }\limits_{#1}}}
\nc{\beq}     {\begin{eqnarray}} 
	\nc{\eeq}    {\end{eqnarray}}
\nc{\nn}      {\\\nonumber} \nc{\vs}      {\vspace{-0.275cm}}
\nc{\fra}    {\frac{1}{2}}
\nc{\mb}        {\mathbf}

\begin{document}
	
	\title{Nonparametric model for the equations of state of neutron star from deep neural network}
	\author{Wenjie Zhou}
	\affiliation{School of Physics, Nankai University, Tianjin 300071,  China} 
	\author[0000-0002-1709-0159]{Jinniu Hu}
	\affiliation{School of Physics, Nankai University, Tianjin 300071,  China} 
	\affiliation{Shenzhen Research Institute of Nankai University, Shenzhen 518083, China} 
	\author[0000-0002-7497-3185]{Ying Zhang}
	\affiliation{Department of Physics, Faculty of Science, Tianjin University, Tianjin 300072, China}
	\author[0000-0003-2717-9939]{Hong Shen}
	\affiliation{School of Physics, Nankai University, Tianjin 300071,  China} 
	
	\email{hujinniu@nankai.edu.cn}
	
	\begin{abstract}
{It is of great interest to understand the equation of state (EOS) of the neutron star (NS), whose core includes highly dense matter. However, there are large uncertainties in the theoretical predictions for the EOS of NS. It is useful to develop a new framework, which is flexible enough to consider the systematic error in theoretical predictions and to use them as a best guess at the same time. We employ a deep neural network to perform a non-parametric fit of the EOS of NS using currently available data. In this framework, the Gaussian process is applied to represent the EOSs and the training set data required to close physical solutions. Our model is constructed under the assumption that the true EOS of NS is a perturbation of the relativistic mean-field model prediction. We fit the EOSs of NS using two different example datasets, which can satisfy the latest constraints from the massive neutron stars, NICER, and the gravitational wave of the binary neutron stars. Given our assumptions, we find that a maximum neutron star mass is $2.38^{+0.15}_{-0.13} M_\odot$ or $2.41^{+0.15}_{-0.14}$ at $95\%$ confidence level from two different example datasets. It implies that the $1.4 M_\odot$ radius is $12.31^{+0.29}_{-0.31}$ km or $12.30^{+0.35}_{-0.37}$ km. These results are consistent with results from previous studies using similar priors. It has demonstrated the recovery of the EOS of NS using a nonparametric model.}
	\end{abstract}
	
	
	\section{Introduction} \label{sec_introd}
	Neutron stars, remnants of very massive stars at the end of their lifecycle,  are one of the most compact objects in the universe, attracting a lot of attention within the fields of astrophysics and nuclear physics~\citep{oertel17}. Rapid developments in space observation technologies and gravitation-wave detection have proven advantageous to the measurement of neutron star properties, such as mass, radius, and tidal deformability. Three massive neutron stars with masses of around $2M_\odot$, PSR J1614-2230~\citep{demorest10,fonseca16,arzoumanian18}, PSR J0348+0432~\citep{antoniadis13}, and  PSR J0740+6620~\citep{cromartie20} have been discovered in the past decade. The gravitational wave from a binary neutron star merger, the GW170817 event, was first detected by the LIGO and Virgo collaborations in 2017, providing a constraint on the tidal deformability of a neutron star at $1.4M_\odot$~\citep{abbott17,abbott18,abbott19}. Furthermore, simultaneous measurements of the mass and radius of PSR J0030+0451 and PSR J0740+6620 were recently analyzed by the Neutron Star Interior Composition Explorer (NICER)~\citep{riley19,miller19,riley21,miller21}.
	
	These studies have improved our knowledge of neutron stars, providing insight into their interior structure and components. A neutron star can be divided into the atmosphere, outer crust, inner crust, outer core, and inner core regions. Its properties are strongly dependent on the equation of state (EOS) of dense nuclear matter~\citep{lattimer00,glendenning01,weber05,lattimer07,baym18}. In the core region, the density approaches $5$-$10$ times the nuclear saturation density. Therefore, in this high-density region, the EOS plays an essential role in investigations, yet it cannot be well determined by current terrestrial methodologies.
	
	Conventionally, the EOS of neutron stars can be extrapolated by the nuclear many-body approaches, such as the density functional theory~\citep{ring96,bender03,meng06,stone07,niksic11,dutra12,dutra14} and {\it ab initio} method~\citep{akmal98,dalen04,sammarruca10,sammarruca12,wei19,wang20}, which can describe the ground-state properties of finite nuclei and nuclear saturation properties very well. However, there are large uncertainties, when these methods are extended to calculate the high-density EOS. They generate many kinds of neutron star mass-radius relations. Furthermore, the isospin dependence of EOS, i.e., the symmetry energy effect, is strongly correlated to the radii of low-mass neutron stars~\citep{li08,bao14,bao15,sun16,ji19,li19,hu20}. With present observations of neutron stars, a smaller slope of symmetry energy, $L$ is preferred. Meanwhile, several exotic hadronic degrees of freedom and/or hadron-quark transitions may appear in the core region of a neutron star because of the phase diagram of strong interaction~\citep{yang08,xu10,chen13,orsaria14,wu17,ju21,huang22}. Hence, it is very difficult to generate a unified EOS in a self-consistent theoretical framework.
	
	Recently, the data-driven methodologies, such as Bayesian inference~\citep{ozel10,raithel17,steiner10,alvarez16,miao21}, deep neural network (DNN)~\citep{fujimoto18,fujimoto20,fujimoto21,farrell22,ferreira22}, nonparametric EOS representation~\citep{landry19,essick20a,essick20b}, support vector machines~\citep{murarka22,ferreira21}, and so on, have been introduced to generate the possible EOSs using the latest observables of neutron stars. In Bayesian inference, the EOS is parameterized and the corresponding parameters are obtained with a marginal likelihood estimation on the posterior probability in terms of model parameters~\citep{ozel10}. Fujimoto {et al.} proposed a scheme that can map the finite observation data of neutron stars onto the EOS with a feed-forward DNN. They present the EOS as a polytropic function with different speeds of sound at distinct density segments~\citep{fujimoto18}. To avoid the limitations of a parametric EOS, Landry {\it et al.} developed a nonparametric method to generate the EOS from the observables of gravitation waves by combining the Gaussian process and Bayesian inference methods, where the EOS of the neutron star is represented by the Gaussian process with finite points~\citep{landry19}. The matching between the EOS and neutron star observations was carried out by the Bayesian inference. Han {et al.} also reconstructed the EOS of a neutron star using another Bayesian nonparametric inference method where the EOS was produced by the neural network with a sigmoid type as the activation function~\citep{han21}.  
	
	In this work, a new machine learning methodology is proposed to reconstruct a  {nonparametric model for the  EOSs of neutron stars} based on the scheme proposed by Fujimoto {et al.}, where the complete EOS is generated by the Gaussian process regression method with finite data points about the pressure-energy relation. A DNN is trained with the constraints of neutron star mass-radius relations, the masses of the heavy neutron stars, and the measurements of NICER. In Section~\ref{2}, the framework of the Gaussian process regression method and the construction of the DNN is given in detail. The {nonparametric  EOS model} of neutron stars generated by the DNN is shown in Section~\ref{3}. A summary is presented in Section~\ref{sum}.
	
	\section{Gaussian process regression and neural network}\label{2}
	\subsection{Gaussian Process Regression}\label{2.1}
	The Gaussian process (GP)~\citep{huang22,williams06}, a random process, is a series of normal distributions of random variables in an index set combination. If the set of random variables $\left\{f(x):x\in \chi \right\}$ is taken from the GP with the mean function $m(x)$ and the covariance function $k(x_1,x_2)$, the corresponding random variables $f(x_i)$ satisfy the multivariate Gaussian distribution for any finite set, $[x_1, \cdots, x_m] \in \chi $, 
	\begin{equation}
		\begin{aligned}
			&\left[\begin{array}{c}
				f(x_1) \\ \vdots \\ f(x_m)
			\end{array}\right]
			\sim 
			&\mathcal{N}\left( 
			\left[\begin{array}{c} m(x_1) \\ \vdots \\ m(x_m) \end{array} \right], 
			\left[\begin{array}{ccc}
				k(x_1,x_1) & \cdots & k(x_1,x_m) \\
				\vdots & \ddots & \vdots \\
				k(x_m,x_1) & \cdots & k(x_m,x_m)
			\end{array}\right]
			\right),
		\end{aligned}
		\label{eq:1}
	\end{equation}
	
	which can be simply expressed as
	\begin{equation}
		f(\cdot) \sim GP(m(\cdot),k(\cdot,\cdot)).
	\end{equation}
	All linear combinations of random variables in GP obey the normal distribution. For each finite-dimensional set, its probability density function on the continuous exponential set is the Gaussian measurement of all random variables. Therefore, it is regarded that the infinite-dimensional set can be generalized by the extension of the multivariate Gaussian distribution. 
	
	Hence, the GP can be applied to solve a normal regression problem,
	\begin{equation}
		y^{(i)} = f(x^{(i)})+\epsilon^{(i)},
	\end{equation}
	where $X$ is defined as the training set and its components $(x^{(1)},...,x^{(m)})$ are independently and identically distributed with unknown distribution. $\epsilon^{(i)}$ is an independent noise variable, which is also given by a normal distribution with variance $\sigma^2$, $N(0,\sigma^2)$. This scheme is called the Gaussian process regression (GPR) method. Usually, it is assumed that $f$ follows the GP with a mean value of zero for notation simplicity,
	\begin{equation}
		f(\cdot) \sim GP(0,k(\cdot,\cdot)).
	\end{equation}
	The test set $X^*=(x^{(1*)},...,x^{(m*)})$, has the same independent co-distribution as $X$, marked as $X\rightarrow X^*$. Therefore, the posterior distribution $p(y^*|X,~X^*)$ is predicted in GPR as the Gaussian distribution of the results, which is different from the general linear regression.
	
	According to the properties of the GP, a joint distribution of the training and test sets is obtained,
	\begin{equation}
		\left.\left[\begin{array}{c} \Vec{f} \\ \Vec{f^*} \end{array}\right]
		\right| X,X^* \sim \mathcal{N}\left(\Vec{0},
		\left[\begin{array}{cc}
			K(X,X)   & K(X,X^*) \\
			K(X^*,X) & K(X^*,X^*)
		\end{array}\right]
		\right)
	\end{equation}
	where the matrix elements $K(X^A,X^B)_{i,j} = k(x^A_i,x^B_j)$. In the GP, the covariance function $k_{ij}$ is also called the kernel function. The standard choice is the squared-exponential kernel,
	\begin{equation}
		k_{se}(x_1,x_2) = \sigma^2 \exp\left(-\frac{||x_1 - x_2||^2}{2l^2}\right).
	\end{equation}
	Meanwhile, their noises obey similar distributions,
	\begin{equation}
		\left[ \begin{array}{c}\Vec{\epsilon} \\ \Vec{\epsilon^*} \end{array}\right]
		\sim \mathcal{N}\left(\Vec{0},
		\left[\begin{array}{cc}
			\sigma^2_{wn} I & \Vec{0} \\
			\Vec{0}^T  & \sigma^2_{wn} I
		\end{array}\right]
		\right).
	\end{equation}
	Here, $\sigma^2_{wn}$ is the hyper-parameter corresponding to white noise, which is different from the signal variance parameter, $\sigma$ in $k_{se}$. The summation of two independent multivariate Gaussian variables is still a multivariate Gaussian variable,
	\begin{equation}
		\begin{aligned}
			&\left.\left[\begin{array}{c} \Vec{y} \\ \Vec{y^*} \end{array}\right]
			\right| X,X^* 
			= \left[\begin{array}{c} \Vec{\epsilon} \\ \Vec{\epsilon^*} \end{array}\right]
			+ \left[\begin{array}{c} \Vec{\epsilon} \\ \Vec{\epsilon^*} \end{array}\right]
			\sim \\ 
			&\mathcal{N}\left(\Vec{0},
			\left[\begin{array}{cc} 
				K(X,X)+\sigma^2_{wn} I   & K(X,X^*) \\
				K(X^*,X) & K(X^*,X^*)+\sigma^2_{wn} I
			\end{array}\right]
			\right)
		\end{aligned}
	\end{equation}
	
	Based on the properties of multivariate Gaussian distribution, the conditional distribution over the unknown $y^*$ is,
	\begin{equation}
		y^*|y,~X,~X^*\sim \mathcal{N}\left(\mu^*,\Sigma^* \right),
	\end{equation}
	where, 
	\begin{equation}
		\begin{aligned}
			\mu^* =\ & K(X^*,X)(K(X,X)+\sigma^2_{wn} I)^{-1}\Vec{y}, \\
			\Sigma^* =\ & K(X^*,X^*) - K(X^*,X) \\
			&(K(X,X)+\sigma^2_{wn} I)^{-1} K(X,X^*).
		\end{aligned}
		\label{eq:10}
	\end{equation}
	$\mu^*$ and $\Sigma^*$ are the mean and covariance functions of the probability distribution for our prediction results, respectively. Therefore, given the hyper-parameters $\sigma$ and $l$ in the kernel function, a probability distribution describing the whole test set by the GPR method can be obtained. In principle, the mean function should be selected as the ``actual data curve". However, it is strongly dependent on the hyper-parameters, $\sigma$ and $l$ that are determined by maximizing the marginal log-likelihood, defined as, 
	\begin{equation}\label{marg}
		\begin{aligned}
			\log p(\bm{y}|\sigma,l) 
			=\ & \log \mathcal{N}(0,K_{yy}(\sigma,l)) \\
			=\ & -\frac{1}{2}\bm{y}^T K_{yy}^{-1} \bm{y} - \frac{1}{2}\log |K_{yy}| - \frac{N}{2}\log(2\pi),        
		\end{aligned}
	\end{equation}
	where $K_{yy}=K(X^*,X^*)$. Therefore, with a small number of data points, a relatively reasonable EOS curve and its confidence range can be predicted in the framework of the GPR method.
	
	The direct matching between the EOS of a neutron star, i.e., the pressure-energy relation, and the observables of a neutron star may generate nonphysical solutions, such as the speed of sound of neutron star matter being less than zero or larger than the speed of light, $c_s < 0$ or $c_s > c$, or the energy density becoming less than zero in some extreme conditions.
	
	Recently, a new intermediate variable $\phi$ was proposed to construct the EOS of a neutron star~\citep{lindblom10,landry19}. $\phi$ is defined as,
	\begin{equation}
		\phi = \mathbf{log}\left(c^2 \frac{d\epsilon}{dp} -1\right).
		\label{eq:12}
	\end{equation}
	It avoids the aforementioned weird behaviors, as when $\phi \in \mathbf{R}$, the speed of sound obeys $0\leq c^2_s=dp/d\epsilon\leq c^2$, which automatically satisfies the physical requirements. When $p>0$, the $\epsilon>0$ can be kept. Due to the large pressure magnitude of pressure, the $\phi$ is regarded as a function of $\log p$ so that it is easier to determine the hyper-parameters. Therefore, Eq.~(\ref{eq:12}) will be expressed as,
	\begin{equation}
		\phi= \mathbf{log}\left(\partial \mathbf{log}\epsilon\frac{ e^{\mathbf{log}\epsilon}}{p}c^2 -1\right),
		\label{eq:13}
	\end{equation}
	where $\partial \mathbf{log}\epsilon=\left.\frac{\partial \log \epsilon}{\partial \log p}\right|_{p = p_i}$. 
	
	In the training set, $n$ data points $(\phi_i,\log p_i)$ are randomly chosen. Once the optimal hyper-parameters are obtained by GPR, the continuum $\phi-\log p$ curve can be generated. The corresponding EOS of the neutron star, $\epsilon(p)$ is provided by numerically integrating
	\begin{equation} \label{ep}
		\frac{\partial \epsilon}{\partial p} = \frac{1+e^{\phi}}{c^2}.
	\end{equation} 
	
	\subsection{DNN method}
	In the available investigations of the structure of neutron stars, the EOS of neutron star matter was first calculated by either the nuclear many-body method or the parameterization function under the conditions of $\beta$-equilibrium and charge neutrality. The EOS was then input to the Tolman–Oppenheimer–Volkoff (TOV) equation~\citep{tolman39,oppenheimer39}, which describes a spherically symmetric and isotropic star in a static gravitational field with general relativity. 
	\begin{equation}
		\begin{aligned}
			\frac{dp}{dr} =\ & 
			-\frac{G\epsilon(r)m(r)}{c^2r^2}
			\left[1+\frac{p(r)}{\epsilon(r)}\right] \\
			&\times\left[1+\frac{4\pi r^3p(r)}{m(r)c^2}\right]
			\left[1-\frac{2Gm(r)}{c^2r}\right]^{-1}\\
			\frac{dm}{dr} =\ & \frac{4\pi r^2\epsilon(r)}{c^2},
			\label{eq:15}
		\end{aligned}
	\end{equation}
	where $r$ is the radial coordinate, representing the distance to the center of the star. The functions $p(r)$ and $\epsilon(r)$ are pressure and energy density (i.e., mass density), respectively. We can easily integrate these differential equations starting at $r=0$, with the initial condition $p(r=0)=p_c$. When it is integrated into the surface of the neutron star, i.e., the radius $R$ and $p(r=R)=0$, then $M=m(R)$ corresponds to the total mass of the neutron star. Therefore, a continuum mass-radius ($M$-$R$) relation of a neutron star can be generated by the TOV equation.
	
	A functional mapping between the EOS space and $M$-$R$ space is constructed through the above framework, in a process called ``TOV mapping". In principle, such mapping is invertible; thus, there should be a relevant inverse mapping~\citep{lindblom92}, where the EOS can be uniquely reconstructed from the observed $M$-$R$ relationship of the neutron star. However, in actuality, the complete $M$-$R$ curve cannot be directly obtained from the observed data due to the discontinuities and uncertainties inherent in neutron star observations~\citep{fujimoto21}. Therefore, a more likely EOS can be inferred from the neutron star observations with uncertainties.
	
	The DNN is a powerful machine learning method to connect the EOS with observed data, following the idea of Fujimoto {et al.}~\citep{fujimoto21}. The neural network (NN) is a representation of the fitting parameters of a function. Deep learning, e.g., the machine learning method using a DNN, is a process of optimizing the parameters contained in the function represented by an NN. Deep learning can be divided into supervised learning and unsupervised learning. The supervised learning that we adopted needs to have specific inputs and outputs before it can complete the fitting process with the training data (i.e., regression).
	
	Compared with general fitting methods, the advantage of deep learning lies in the generalization properties of NNs. It does not need to rely on any prior knowledge about the proper form of the fitting function. Due to a large number of neurons (and neuron layers) and fitting parameters, an NN with a sufficient number of neurons can generate any continuous
	function~\citep{cybenko89,hornik91}.
	
	The model function of a feed-forward NN can be expressed as,
	\begin{equation}
		\begin{aligned}
			\bm{y} = f(\bm{x}|\{&W^{(1)},b^{(1)},\cdots,W^{(l)},b^{(l)},\cdots, \nn
			&W^{(L)},b^{(L)}\})
		\end{aligned}
	\end{equation}
	where $\boldmath{x}$ and $\boldmath{y}$ are the inputs and outputs, respectively.
	$W^{(l)}$ and $b^{(l)}$ represent the weights of the middle layer and are given in matrix and vector form respectively. The calculation process of each layer of neurons is,
	\begin{eqnarray}
		\bm{x}^{(0)}&=& \bm{x} \\\nonumber
		\bm{x}^{(l)} &=& \sigma^{(l)}(W^{(l)}\bm{x}^{(l-1)}+b^{(l)}),~~~~~~(l = 1,\cdots,L)
	\end{eqnarray}
	The $L$-th layer is the output one, $\bm{y}=f(\bm{x})=\bm{x}^{(L)}$. Here $\sigma^{(l)}(x)$ is called the activation function, which can make the relationship between neurons of each layer not only be linear but also increase the complexity of the NN. A typical activation function has a rectified linear unit ($\sigma(x)=\max\{0,x\}$), a hyperbolic tangent ($\sigma(x)=\tanh(x)$), a sigmoid function ($\sigma(x)=1/(e^x+1)$), and so on.
	
\begin{figure}[htbp]
		\centering
		\includegraphics[width=0.8\linewidth]{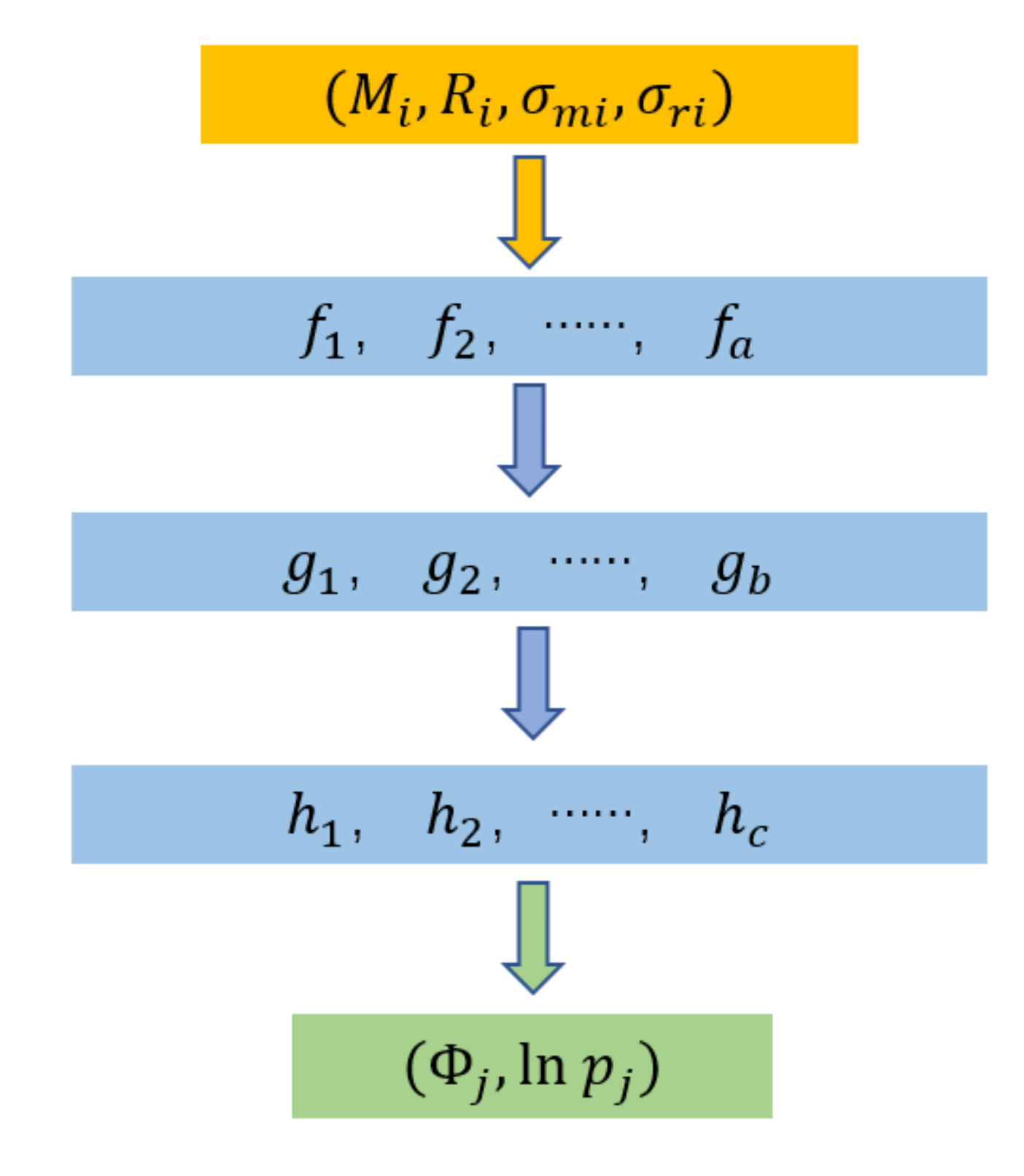}
		\caption{The NN flow chart of present framework.}
		\label{fig1}
	\end{figure}

	When the number of neuron layers, the number of neurons ($a,~b,~c$), and the corresponding activation function ($f,~g,~h$) are fixed, a basic NN is built, as shown in Fig.~\ref{fig1}. Here, $M$-$R$ observation data was selected as the input layer, and the variable $\phi_i$, corresponding to $p_i$, was set up as the output layer, which is the reverse process when compared with other studies on neutron stars.
	
	To optimize the NN to generate the best result during training, we also need to define a loss probability to evaluate the training results, which is written as,
	\begin{equation}
		\begin{aligned}
			\mathcal{L}\left(\left\{W^{(l)},b^{(l)}\right\}_l\right) \equiv \int 
			&d\bm{x} Pr(\bm{x}) \ell\left(\bm{y},f\left(\bm{x}|
			\left\{W^{(l)},b^{(l)}\right\}_l\right)\right).
		\end{aligned}
		\label{eq:18}
	\end{equation}
	
	Here, $\ell(\bm{y},\bm{y}')$ quantifies the distance or error between the predicted $\bm{y}'$ of NN and the result $\bm{y}$ from the training data. The small-batch method was used to evaluate its derivatives, where the training data set $\mathcal{D}$ is first randomly divided into multiple subsets. Then, the derivative of the loss probability is estimated in a small batch of $\mathcal{B}$, and the approximate derivative is,
	\begin{equation}
		\frac{\partial\mathcal{L}(W^{(l)})}{\partial W^{(l)}} \approx \frac{1}{|\mathcal{B}|}  \sum_{n=1}^{|\mathcal{B}|}
		\frac{\partial\ell \left(\bm{y}_n,f\left(\bm{x}_n|
			\left\{W^{(l)},b^{(l)}\right\}_l\right)\right)}
		{\partial W^{(l)}},
		\label{eq:19}
	\end{equation}
	where, the batch size $|\mathcal{B}|$ represents the number of sample points in $\mathcal{B}$. Since each optimal choice varies from case to case, its error will be shown later as a part of our estimations on the EOS confidence. The epoch denotes the number of scans of the entire training data set $\mathcal{D}$. Parameters are updated with each small batch, so an epoch is equivalent to iterating $|\mathcal{D}|/|\mathcal{B}|$ small batches of data until all iterations are completed. In addition, the derivative $\frac{\partial\ell}{\partial W}$ that appeared in Eq.~(\ref{eq:19}) was calculated by the back-propagation method.
	
	In this training process, the mean square logarithmic error (msle) is regarded as the loss $\ell(\bm{y},\bm{y}')$ in Eq.~(\ref{eq:18}),
	\begin{equation}~\label{msle}
		\ell_{\rm msle}(\bm{y},\bm{y}') \equiv |\log \bm{y} - \log \bm{y}'|^2.
	\end{equation}
	With a loss function, our NN can begin the basic training. The parameter initialization of NN will be discussed later, in detail.
	
    {Therefore, it is useful to compare our method with other methods proposed to generate the EOS of neutron star.}	In the present framework, the {fitted} EOSs are obtained by the DNN. The neutron star observation data is chosen as the input layer, while the constraint EOS is set up as the output layer. The training process is finished with the observations' likelihoods and the EOS priors generated by the theoretical model. The EOSs in the priors and the output layer are presented by several discretized points in {$\phi$}-function to satisfy the constraint of the speed of sound and are smoothly connected by GP. On the other hand, the EOSs in the work of Fujimoto et al. were parameterized as a polytrope function dependent on the speeds of the sound of neutron star matter. Furthermore, the {fitted} EOSs in the work of Landry and Essick were produced by Bayesian inference with a set of nuclear-theoretic models.
	
	\section{The numerical details and results}\label{3}
	To prepare the training data set, the EOSs from relativistic mean-field (RMF) models were used to obtain the generation interval of GPR fitting data points. Nine RMF parameterizations were selected: BigApple, DD2, DDLZ1, DDME1, DDME2, DDMEX, NL3, PKDD, and TW99~\citep{bigapple,dd2,ddlz1,ddme1,ddme2,ddmex,nl3,pkdd}. All of these RMF parameter sets can provide neutron stars, whose maximum masses are larger than $2.0M_{\odot}$~\citep{huang20}. The EOS from the NL3 set generated a maximum mass of neutrons star around $2.78M_{\odot}$.
	
\begin{figure}[htbp]
		\centering
		\includegraphics[width=0.9\linewidth]{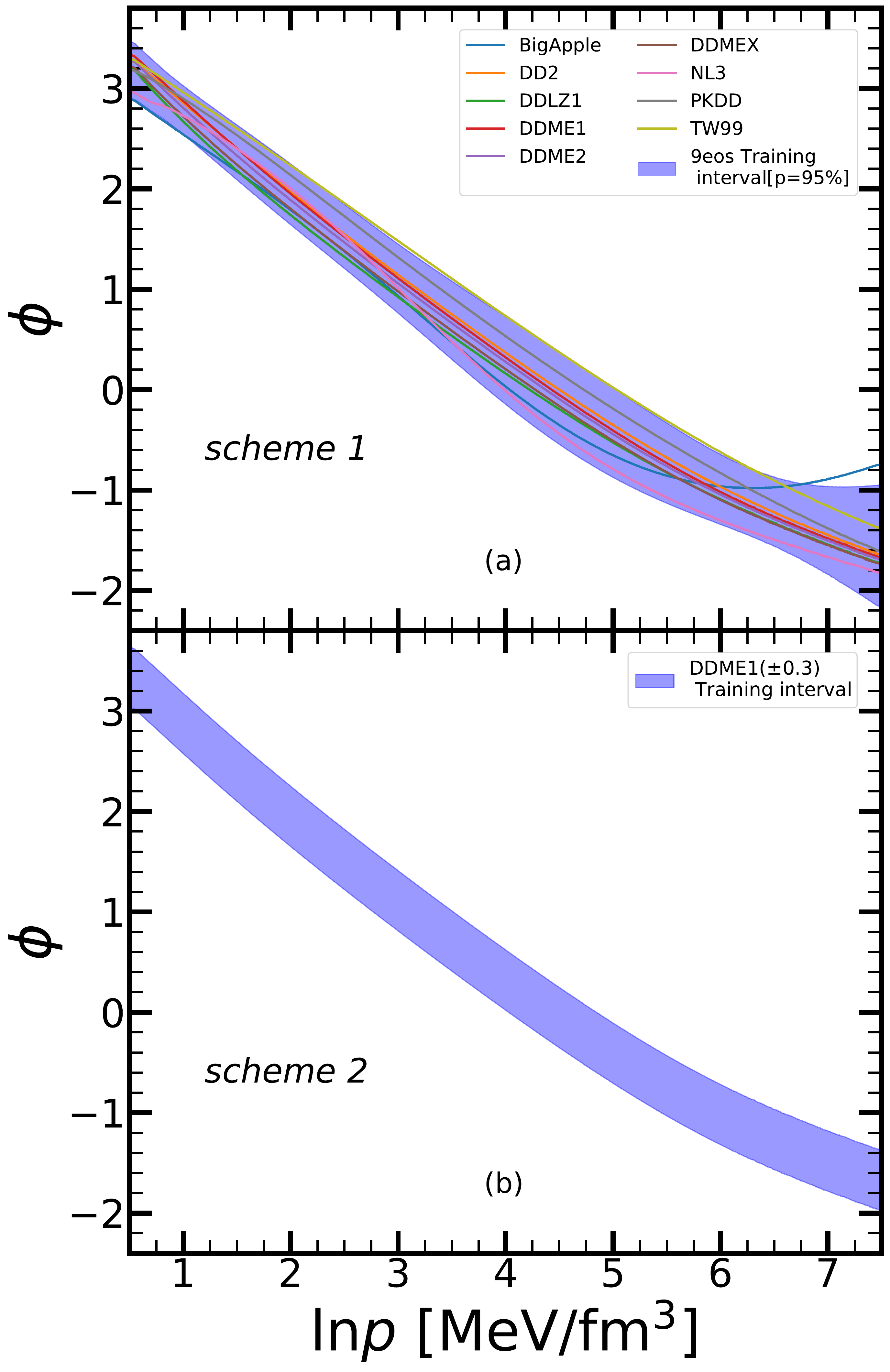}
		\caption{The generation range of $\phi$-$\ln p$. We will randomly select points within this range and then use GPR method to generate EOS. In panel (a ) the nine EOSs are treated to obtain the mean $\mu$ and variance $\sigma$, whose $95\%$ confidence interval is taken to obtain the fitting range. In panel (b), the generation range is based on DDME1 curve, with a fluctuation of $0.3$.}
		\label{fig2}
	\end{figure}

	The $\epsilon$-$p$ relation in the EOS was transferred into the $\phi$-$\ln p$ function, where $\ln p$ is the natural logarithm of pressure. After calculating the means and variances of the $\phi$-$\ln p$ relations from the above nine EOSs, it was found that their mean value is very close to the EOS from the DDME1 set~\citep{ddme1}. To investigate the stability of initial values in the present framework, two schemes were adopted to generate the fitting interval with the GPR method:
	\begin{enumerate}
		\item [1)] $\emph{Scheme\ 1}$ -- After obtaining the mean and variance of $\phi$-$\ln p$ functions from nine RMF parameter sets, the $95\%$ confidence interval of the variance was selected as the generation range of $\phi_i$. As shown in panel (a) of Fig.~\ref{fig2}, this interval encloses all EOSs from the RMF model.
		
		\item [2)] $\emph{Scheme\ 2}$ -- The $\phi$-$\ln p$ function provided by DDME1 set was regarded as the
		standard, and $\phi\pm 0.3\phi$ are chosen as the upper and lower bounds of the generation range of $\phi_i$. Such an interval is consistent with the one obtained by scheme 1, to a large extent.
	\end{enumerate}

\begin{figure}[htbp]
	\centering
	\includegraphics[width=0.9\linewidth]{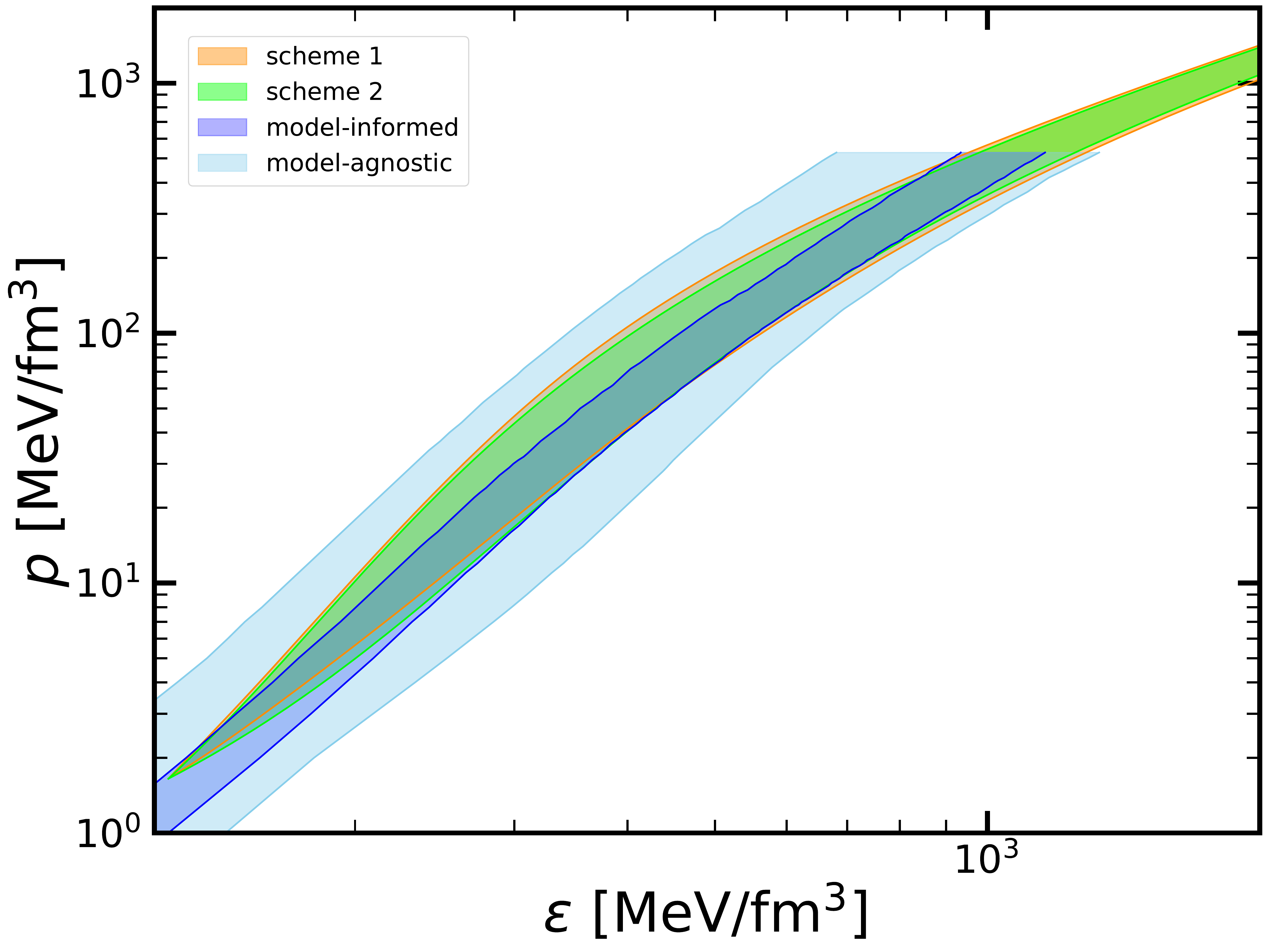}
	\caption{The corresponding $\epsilon-p$ relations of scheme 1 and 2 in Figure 2 and the model-informed and model-agnostic priors in the Bayesian inference method by Landry and Essick~\citep{landry19}.}
	\label{fig21}
\end{figure}

In Fig.~\ref{fig21}, the corresponding $\epsilon$-$p$ relations of scheme 1 and 2 are compared to the model-informed and model-agnostic priors in the Bayesian inference method by Landry and Essick~\citep{landry19}. The $\epsilon$-$p$ relations from scheme 1 and scheme 2 in the present work are almost identical, which are also consistent with the model-informed prior. Since all of them are more strictly constrained by the theoretical EOSs. On the contrary, the model-agnostic prior has a loose boundary. It may consider more range of plausible EOSs.

To produce an EOS of neutron stars (including the high-density region) with the GPR method and aforementioned schemes, seven pressure points $\ln p_i~(i=1,~2,~\cdots,~7)$ were selected, with the same interval, in the range $\ln p\in [1,7]$. $\phi_i$ was randomly generated in the training interval at each $\ln p_i$ point as an initial data set $(\phi_i,~\ln p_i)$. The EOS below nuclear saturation density was chosen as the one from the SLy4 set. A smooth and continuous $\phi(\ln p)$ function is {fitted} by the GPR method, where the hyper-parameters, $l$ and $\sigma$ are obtained by maximizing the marginal log-likelihood, as shown in Eq.~(\ref{marg}). Furthermore, the star point, $\phi_1=\phi(\ln p=1)$ was fixed as the magnitude from the DDME1 parameter set.
	
The $M$-$R$ relation of a neutron star can be calculated using the EOS from the GPR method by solving the TOV equation. In the present framework, the training data set of the DNN should assemble the points on the  $M$-$R$ curve, which correspond to the observables. The method proposed by  Fujimoto {et al.}~\citep{fujimoto21}  is used in this work to generate training data. Firstly, the maximum masses of neutron stars less than $2.2M_\odot$ and the $M$-$R$ relations that did not satisfy the radii constraints of PSR J0740+6620 and PSR J0030+0451~\citep{miller19,miller21} were excluded from the training data. Then, $14$ points in the mass regions, $\left[M_{\odot}, M_{max}\right]$  on the $M$-$R$ curve were randomly chosen as ``the original data points''  $(M_i, R_i)$ to simulate the real observations of the $14$ available neutron stars. To consider the errors in the observations, the variances of the Gaussian distributions about the mass and radius, $\sigma_{M_i}$ and $\sigma_{R_i}$, were randomly taken from the uniform distribution in the ranges, $\left[0,M_{\odot}\right]$ and $\left[0,5\textup{km}\right]$. The deviations of mass and radius $\left(\Delta M_i, \Delta R_i\right)$ were calculated  by the Gaussian distribution with the variances of $\sigma_{M_i}$ and $\sigma_{R_i}$. Finally the ``real data point" $\left(M_i+\Delta M_i, R_i+\Delta R_i\right)$ was obtained. The set $\left(M_i+\Delta M_i,~R_i+\Delta R_i,~\sigma_{M_i},~\sigma_{R_i}\right)$ can be compared to the observational data of neutron stars.
	
	A group of $i=14$ data points $(M_i,~R_i)$ was selected from the $M$-$R$ curve generated by each EOS, and  $j=100$ groups of different variances $(\sigma_{M_{ij}},\sigma_{R_{ij}})$ were randomly sampled for each $M_i$-$R_i$ data point. Later, $k=100$ groups of deviations, $\Delta M_{ijk}$ and $\Delta R_{ijk}$ were provided by each variance set, $(\sigma_{M_{ij}},\sigma_{R_{ij}})$. 
	In this way, $100\times100$ sets of data for each EOS were prepared and $14$ data points were sampled. The above process was repeated by $500$ times to include as wide a range as possible, resulting in $500\times100\times100= 5,000,000$ sets, where one set includes $14$ data points. 
	
	\begin{table*}[htb]
		\centering
		\begin{tabular}{c|c|c} 
			\hline \hline
			Layer & Number of neurons & Activation function \\\hline
			1(Input) & 56 & N/A \\\hline 
			2 & 60 & ReLU \\\hline
			3 & 40 & ReLU \\\hline
			4 & 40 & ReLU \\\hline
			5(Output) & 6 & tanh \\\hline \hline
		\end{tabular}
		\caption{The setup of present  DNN. The number of input and output neurons can be modified according to different network conditions. Here, the number of neurons at output layer is $6$, because $\phi(\ln p =1)$ has been fixed as the value obtained from DDME1 set.}\label{tab1} 
	\end{table*}
	
	For the architecture of the NN, the Python library, Keras~\citep{chollet15} was employed, with TensorFlow~\citep{abadi16} as the backend. The number of NN layers, their corresponding neurons, and the activation functions are shown in Table~\ref{tab1}. The hyperbolic tangent function of the output layer makes the results fall between $(-1,1)$, speeding up the training.
	The msle is chosen as the loss function, given in Eq.~(\ref{msle}). The optimization method was Adam~\citep{kingma14} by taking the batch size as $1000$. The default initialization NN argument was the Glorot Uniform distribution~\citep{glorot10}.
	
\begin{figure}[htbp]
		\centering
		\includegraphics[width=0.9\linewidth]{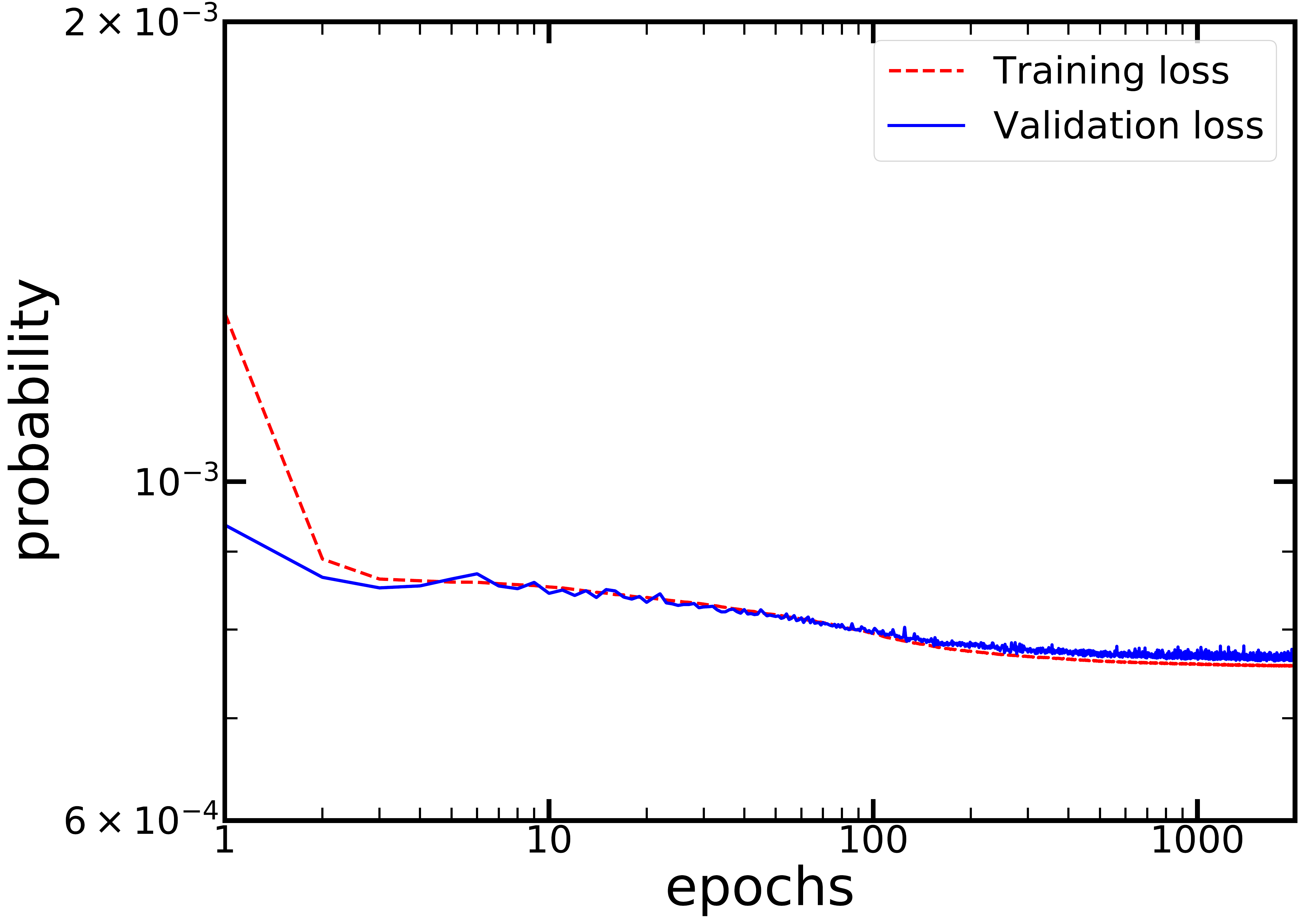}
		\caption{The Loss probabilities as functions of epoch with the training data and validation data.}
		\label{fig3}
	\end{figure}

	The DNN models for a full training set of $5,000,000$ data were compare with a random sampling of $1,000,000$ data in the training set, giving similar results, but with the latter greatly improving the training efficiency. In addition, for all models, the changes in loss functions for the training of epoch were almost identical. The loss functions estimated for the validation data and training data are shown as an example in Fig.~\ref{fig3}. When the epoch $>10$, the verification loss is consistent with the training loss, whereas when the epoch $>100$, the verification loss is stable. Therefore, each DNN model was trained with $1,000,000$ data. The validation set was taken as the $10,000$ sets from the rest $4,000,000$ sets  to check the convergence. Once the epoch $= 100$, the model was considered finished.
	
\begin{figure}[htbp]
		\centering
		\includegraphics[width=0.9\linewidth]{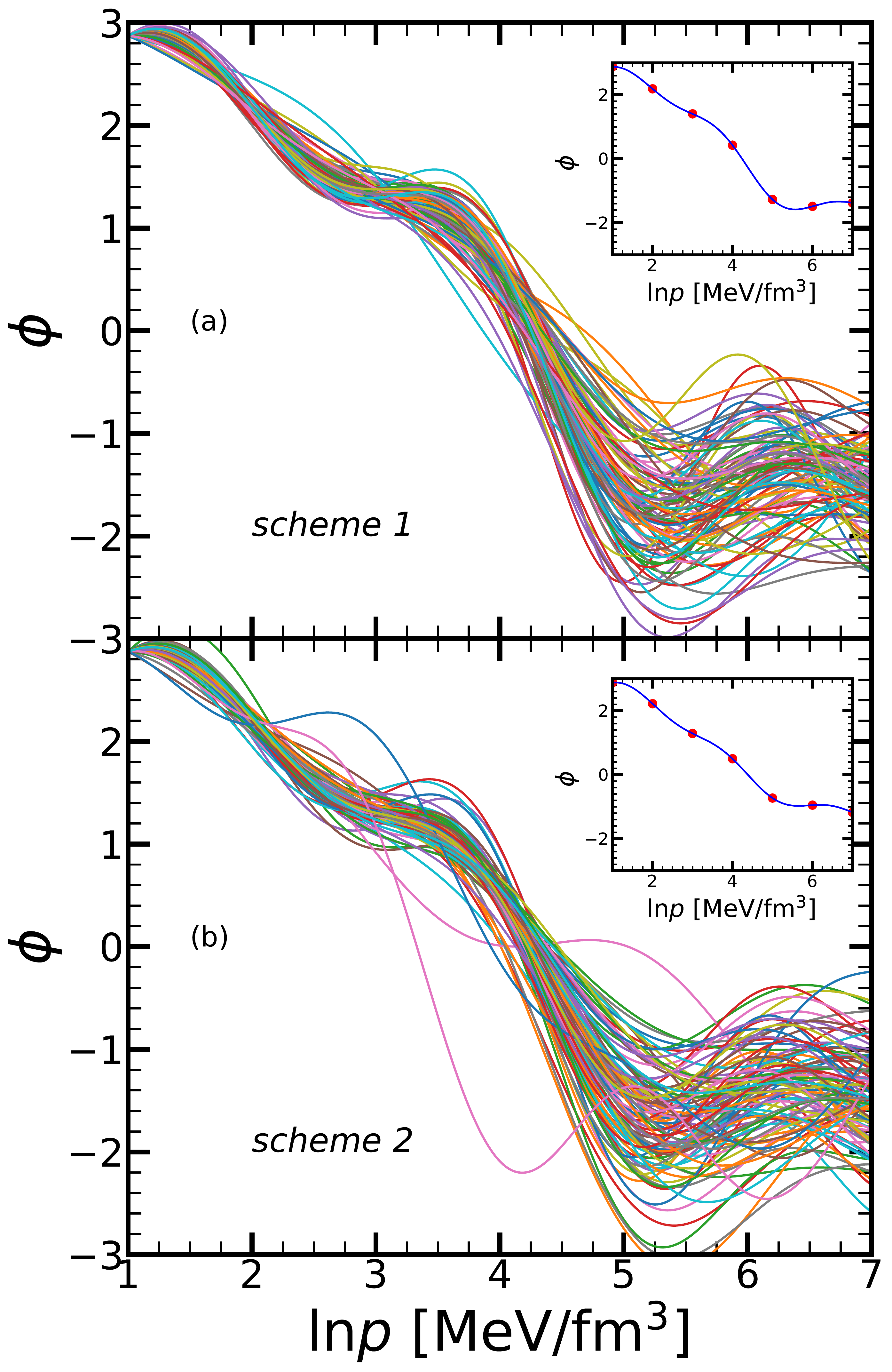}
		\caption{The $200$ DNN models about $\phi$-$\ln p$ from schemes 1 and 2.}
		\label{fig4}
	\end{figure}

	Due to the differences in initial input and training data, there was some uncertainty about the output results of the DNN. Therefore, the process was repeated $100$ times to generate $100$ independent DNN models. The uncertainties in the training results were estimated from the {fitted} $100$ EOSs. In Fig.~\ref{fig4}, $200$ relations about $\phi$-$\ln p$ from scheme 1 in panel (a) and scheme 2 in panel (b) are reconstructed through the training data of the DNN. Each curve is smoothly connected with seven output points by the GPR method, as shown in the inserts. It was found that most of these curves have similar pressure-dependence behaviors. Their differences increase in the high-density region due to the observation discrepancies associated with the $14$ neutron stars.
	
\begin{figure}[htbp]
		\centering
		\includegraphics[width=0.9\linewidth]{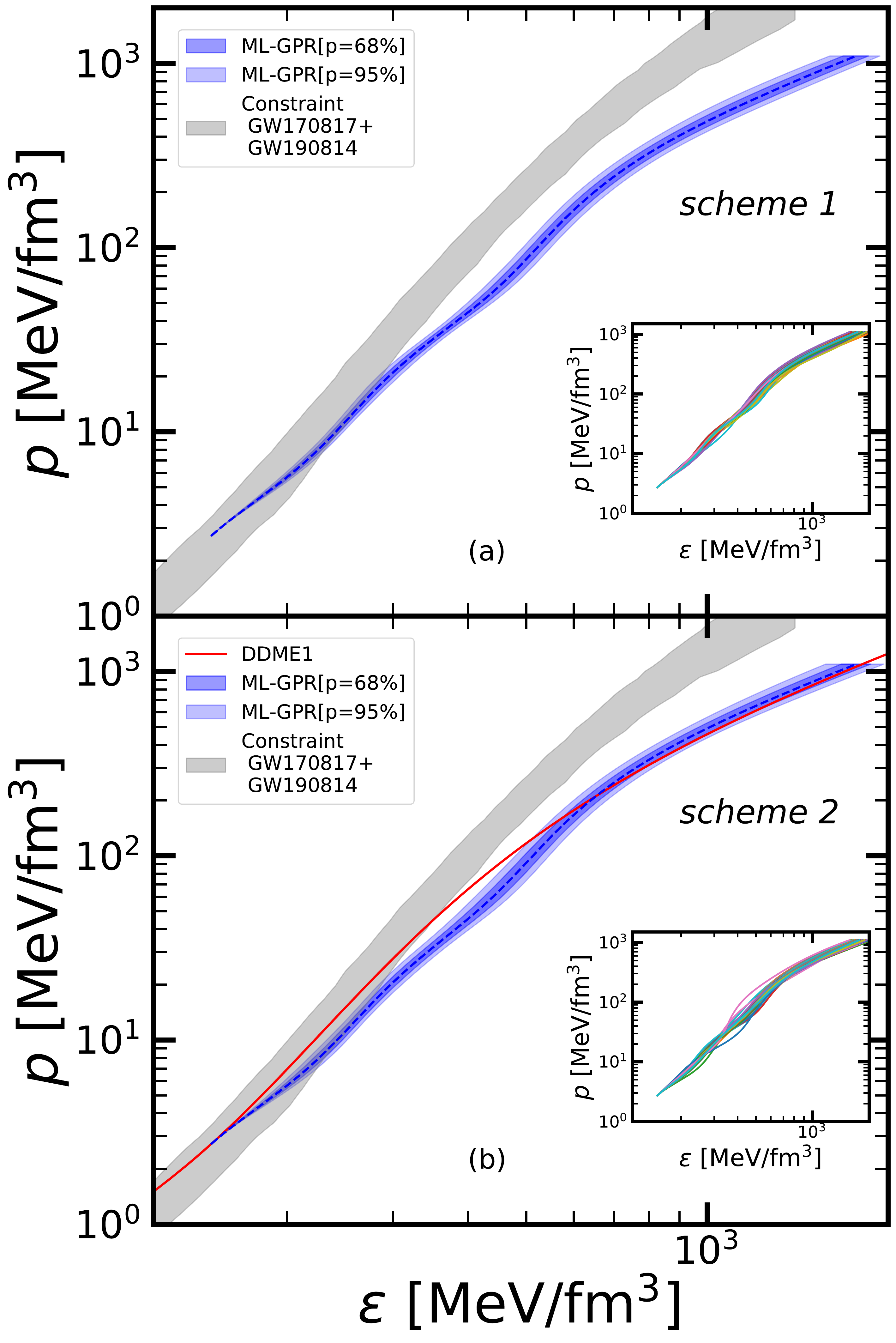}
		\caption{The EOSs from the nonparametric machine learning methods with scheme 1 and 2 and comparing to those from the joint constraints from GW170817 and GW190814 events, and from the DDME1 set.}
		\label{fig5}
	\end{figure}

	The $\phi$-$\ln p$ relations must be converted to the $\epsilon$-$p$ function by integrating the Eq.~(\ref{ep}) to obtain the EOS of the neutron star. In Fig.~\ref{fig5}, the neutron star EOSs with the $68\%$ and $95\%$ confidence levels from the DNN with scheme 1 in panel (a) and scheme 2 in panel (b) are shown and compared to those joint constraints from the GW170817 and GW190814 events~\citep{abbott20b} and the EOS from DDME1. In the inserts, the original $200$ EOSs from the DNN training are plotted. To analyze the uncertainties of the EOSs, it was assumed that the pressures at each energy density from the machine learning model satisfy the Gaussian distribution. Therefore, the mean EOS was obtained as the dashed curve with the dark blue shadow representing the $68\%$ confidence level and the light blue shadow, the $95\%$, respectively. In the low-density region, our estimations are consistent with the joint constraints on the EOS from the GW170817 and GW190814 events. With density increasing, present EOSs are softer than the joint constraints, since the maximum masses of the $14$ neutron stars are just around $2M_\odot$. Furthermore, the {fitted} EOS differs slightly from the EOS of DDME1 in scheme 2, despite this being regarded as the mean value of the training data. In the mediate region of energy density, the EOS generated by the DDME1 is harder than the {fitted} one, since the radius of the neutron star from DDME1 is a little larger when compared with the observations of the $14$ neutron stars, as shown later. These results demonstrate that the EOS of the present framework is independent of the initial input of the training set.
   
Here, it must be emphasized that the inconsistencies in EOSs {fitted} by LIGO-Virgo-KAGRA (LVK) collaborations from GW170817 and GW190814 events, and present work are generated by the different theoretical frameworks and priors. In the LVK analysis, the EOSs in the priors were given by the spectral representation and are determined by the adiabatic index $\Gamma$ as shown in Refs.~\citep{read09} and \citep{lindblom10}. The EOS parameters of the prior ranges in LVK were choices from the 34-neutron star matter EOSs, including the PAL6, APR1-4, WFF1-3, MS1-2, and so on~ \citep{read09}. The maxim masses of the neutron star from these EOSs are in the range of $1.47\sim2.78 M_\odot$ and the radii at $1.4M_\odot$ are $9.36\sim15.47$ km. Correspondingly, the prior of EOSs space in the present framework is taken from the $9$ RMF parameter sets, which only can generate the maximum masses of the neutron stars from $2.0\sim2.4 M_\odot$. Therefore, the harder EOSs were {fitted} by LVK at high-density regions.
	
\begin{figure}[htb]
		\centering
		\includegraphics[width=0.9\linewidth]{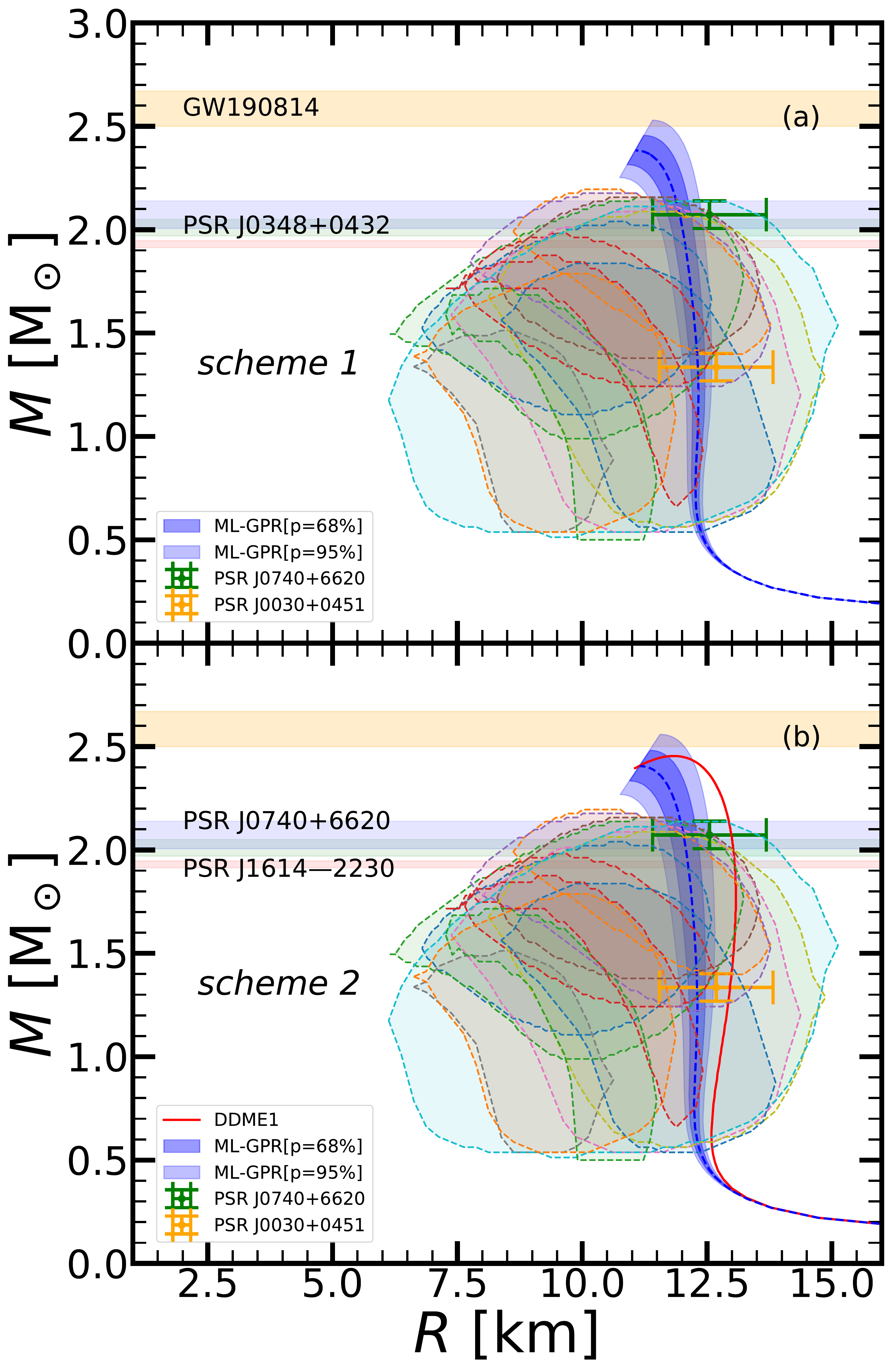}
		\caption{The mass-radius relation of neutron star from the  nonparametric  machine learning method, the observation distributions from $14$ neutron stars, the masses of massive neutron stars, and the radii constraints from the NICER.}
		\label{fig6}
	\end{figure}

	Once the EOS of the neutron star were determined, its $M$-$R$ relation was obtained by solving the TOV equation. The $M$-$R$ relations from our deduced EOSs are plotted in Fig.~\ref{fig6}, with $68\%$ (dark blue) and $95\%$ (light blue) confidence levels. The corresponding $M$-$R$ distributions of the observed $14$ neutron stars are given as contour plots. The masses of massive neutron stars, PSR J0348+0432, PSR J0740+6620, and PSR J1614-2230; the secondary compact object of the GW190814 event; and the radii of PSR J0030+0451 and PSR J0740+6620 from the NICER are given and compared. The {fitted} EOSs from schemes 1 and 2 nicely reproduce the neutron star observations and are able to generate massive neutron stars. Their radii are consistent with the results of the $14$ observed neutron stars and the mass-radius simultaneous measurements from NICER. Furthermore, the $M$-$R$ relation from the DDME1 set is shown as a solid line, which was chosen as the mean value to generate the training data set in scheme 2. Its radius at the mediate mass region is a little larger when compared with the $14$ observed neutron stars. The output EOSs of the DNN from scheme 1 provide smaller radii, which coincide with the distribution of observables. This shows that the final results of present framework is independent of the generating scheme for the training data.
	
\begin{figure}[htb]
		\centering
		\includegraphics[width=0.9\linewidth]{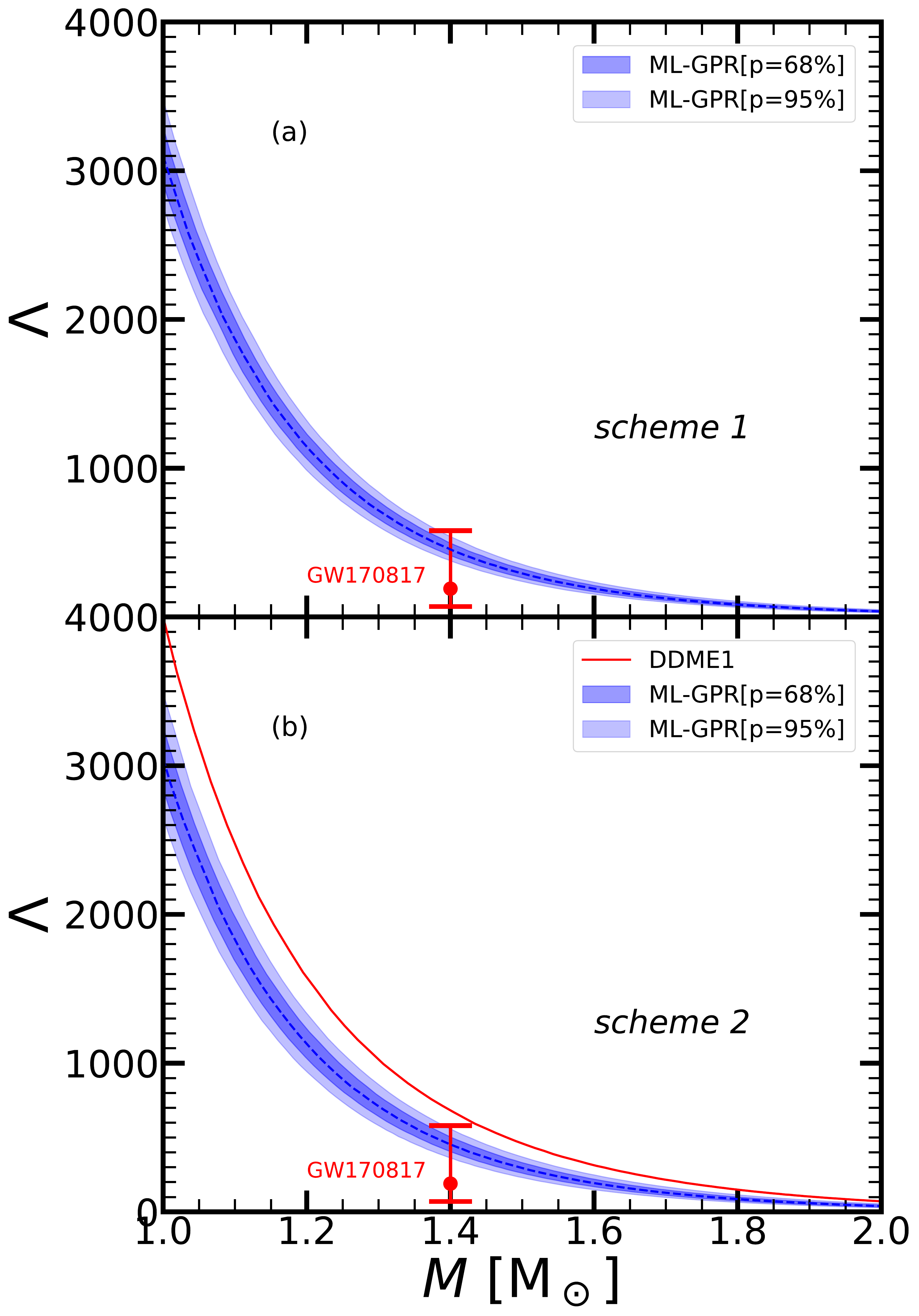}
		\caption{$\Lambda$-$M$ relation, generated by the {fitted} EOSs and compared to that from DDME1 and the values extracted from GW170817 events.}
		\label{fig7} 
	\end{figure}

	In a binary neutron star merger, one neutron star will be deformed by the external gravitational field of another star. The magnitude of deformation is denoted as the tidal deformability, which is dependent on the EOS of the neutron star and can be extracted from the gravitational wave provided by the binary neutron star. In the GW170817 event, the dimensionless tidal deformability at $1.4M_\odot$ was inferred as $\Lambda_{1.4}=190^{+390}_{-120}$~\citep{abbott18}. In Fig.~\ref{fig7}, the dimensionless tidal deformabilities as functions of neutron star masses from schemes 1 and 2,  with $68\%$ and $95\%$ confidence levels, are plotted and compared to the constraint from the GW170817 event and the results from the DDME1 set. The $\Lambda$ decreases with the neutron star mass since it is proportional to $R^5/M^5$ of the neutron star. Therefore, the $\Lambda$ from the DDME1 is relatively larger. The $\Lambda_{1.4}$ from the reported machine learning framework completely satisfies the measurements from the gravitational wave detection.
	
	\begin{table*}[htb]
		\centering
		\begin{tabular}{cccccccc}
			\hline
			&   C. L.     & ~$M_{max} [M_\odot]$ ~&~ $R_{max}$ [km] ~& ~$R_{1.4}$ [km]~& ~$R_{2.08}$ [km]~ & ~$\Lambda_{1.4}$~ \\ \hline
			DDME1   &   & 2.45   & 11.83   & 12.99  & 12.98   & 692 \\ \hline
			\multirow{4}{*}{\begin{tabular}[c]{@{}c@{}}scheme 1\end{tabular}}   & \multirow{2}{*}{$68\%$} & \multirow{2}{*}{$2.38_{-0.07}^{+0.07}$} & \multirow{2}{*}{$11.07_{-0.17}^{+0.16}$} & \multirow{2}{*}{$12.31_{-0.16}^{+0.15}$} & \multirow{2}{*}{$11.95_{-0.23}^{+0.23}$} & \multirow{2}{*}{$459_{-46}^{+37}$} \\
			&  &   &   &   &    &    &   \\ \cline{2-8} 
			
			& \multirow{2}{*}{$95\%$} & \multirow{2}{*}{$2.38_{-0.13}^{+0.15}$} & \multirow{2}{*}{$11.07_{-0.32}^{+0.34}$} & \multirow{2}{*}{$12.31_{-0.31}^{+0.29}$} & \multirow{2}{*}{$11.95_{-0.47}^{+0.44}$} & \multirow{2}{*}{$459_{-81}^{+82}$} \\
			&  &   &   &   &    &    &   \\ \cline{2-8}  \hline

			\multirow{4}{*}{\begin{tabular}[c]{@{}c@{}}scheme 2\end{tabular}}   & \multirow{2}{*}{68$\%$} & \multirow{2}{*}{$2.41_{-0.07}^{+0.08}$} & \multirow{2}{*}{$11.15_{-0.20}^{+0.21}$} & \multirow{2}{*}{$12.30_{-0.19}^{+0.17}$} & \multirow{2}{*}{$12.03_{-0.27}^{+0.27}$} & \multirow{2}{*}{$448_{-43}^{+55}$} \\
			&  &   &   &   &    &    &   \\ \cline{2-8} 
			
			& \multirow{2}{*}{95$\%$} & \multirow{2}{*}{$2.41_{-0.14}^{+0.15}$} & \multirow{2}{*}{$11.15_{-0.39}^{+0.41}$} & \multirow{2}{*}{$12.30_{-0.37}^{+0.35}$} & \multirow{2}{*}{$12.03_{-0.54}^{+0.53}$} & \multirow{2}{*}{$448_{-86}^{+110}$} \\
			&  &   &   &   &    &    &   \\ \cline{2-8}  \hline
		\end{tabular}
		
		\caption{The maximum masses of neutrons star, the corresponding radii, the radii at $1.4 M_\odot$ and $2.08 M_\odot$, and the dimensionless tidal deformability at $1.4 M_\odot$  from the  {nonparametric EOS models} with $68\%$ and $95\%$ confidence levels in scheme 1 and 2 and compared to those from DDME1. }\label{tab2}
	\end{table*}
	
	Table~\ref{tab2} lists the properties of neutron stars {fitted} by the DNN with  nonparametric training data: namely, the maximum masses of neutrons stars, the corresponding radii, the radii at $1.4 M_\odot$ and $2.08 M_\odot$, and the dimensionless tidal deformability at $1.4 M_\odot$ with $68\%$ and $95\%$ confidence levels in schemes 1 and 2. These variables were compared to the results from the DDME1 parameter set. Both of these two schemes can generate the massive neutron star with a mass close to $2.55M_\odot$. The radius of the $1.4M_\odot$ neutron star is around $12.30$ km, which is consistent with the value extracted from the GW170817 of $R_{1.4}=11.9\pm1.4$ km~\citep{abbott19}. The radius of $2.08M_\odot$ neutron star is {fitted} around $12.0$ km now. The radius and mass of PSR J0740+6620 were analyzed as $12.39^{+1.30}_{-0.98}$ km and $2.072^{+0.067}_{-0.066} M_\odot$, from NICER, by Riley {et al.}~\citep{riley21}. The results from the two schemes are similar, with differences are less than $2\%$.
	
	 It can be found that present fits about the properties of the neutron are comparable with those generated by model-informed priors in the works of Landry and Essick~\citep{landry19}, while they are much more constrained than the ones from model-agnostic prior. It is because our training data is just prepared to reproduce the theoretical EOSs, while the possibility that the EOS might be quite different from current theoretical fits was considered in model-agnostic prior.
	
\begin{figure}
		\centering
		\includegraphics[width=0.9\linewidth]{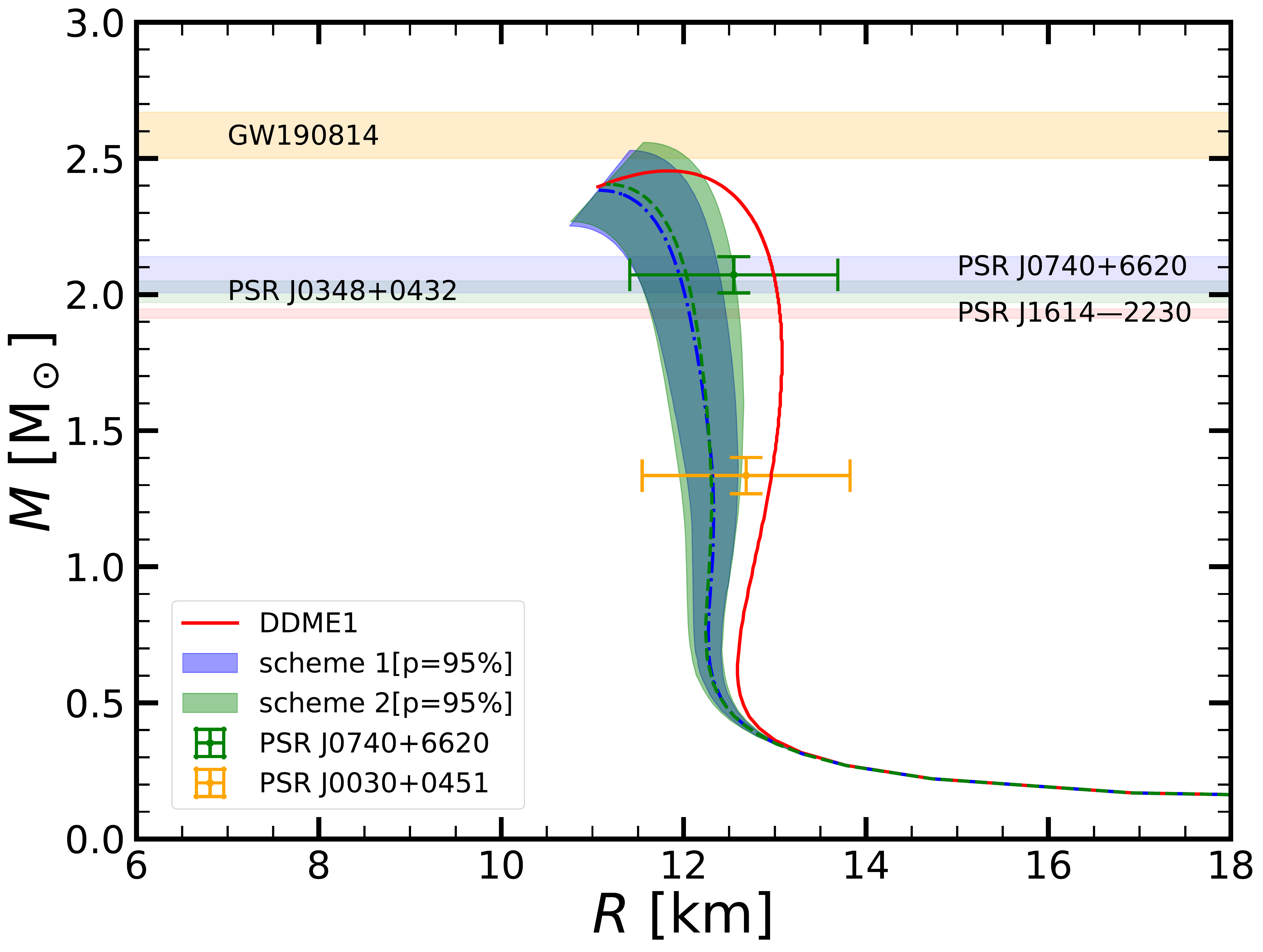}
		\caption{The $M$-$R$ relation comparisons between two schemes with $95\%$ confidence interval and the constraints from the massive neutron star and NICER.}
		\label{fig8}
	\end{figure}

	Finally, the $M$-$R$ relations from the two schemes to generate the training set, were compared and given in Fig.~\ref{fig8}. Their behaviors are quite similar. The only difference is that the radii of the neutron stars and the uncertainties from scheme 2 are a little larger than those of scheme 1 because of the influence of the DDME1 set. This demonstrates that the {fitted} EOSs in the present framework is strongly independent of the choice of initial training data values using the GPR method.
	
	\section{Summaries and perspectives}\label{sum}
	A nonparametric methodology has been proposed to infer the EOSs of neutron star matter from recent observations of the neutron stars. A DNN was designed to map the mass-radius observables to the energy-pressure relation of dense matter. The GPR  method was applied to construct the EOSs, and this method was completely independent of any apparent function form.
	
	To generate the training data set, two schemes of the example data were adopted to provide the initial EOS. The mean values and variances of EOSs from nine successful relativistic mean-field model parameter sets were considered in the first scheme; whereas in the second, the mean value was chosen from the DDME1 set and the derivation was fixed as $0.3$.  A $5$-million training data set was constructed by including the uncertainties in the mass and radius of neutron stars. Furthermore, in the training set, the constraints of the massive neutron star and the mass-radius simultaneous measurements were also taken into account in the training set.
	
	One hundred independent NN models were generated with different training data sets, producing one hundred EOSs of a neutron star. These were analyzed with the standard statistical method and EOSs with the $68\%$ and $95\%$ confidence levels were obtained. They were softer when compared with the join constraints from the GW170817 and GW190814 events. The mass-radius relations from our {fitted} EOSs fully satisfy the present various astronomical observations of neutron stars. The dimensionless tidal deformability at $1.4M_\odot$ was also consistent with the data extracted from the GW170817. Finally, concerning the creation of training data, the results from both schemes were almost identical. This shows that the present {fitted} EOSs are strongly independent of the initial set of training data set.
	
	Our nonparametric NN framework can be naturally extended to other supervised learning fields to avoid the limitations of specific function forms. In the future, the original data on the gravitational wave from the binary neutron star will be included in the input layer to simulate the observations more realistically. The hadron-quark phase transition was excluded in the present training data set, and this too will be considered in future work.
	
	\section{Acknowledgments}
	This work was supported in part by the National Natural Science Foundation of China	 (Grant  Nos. 11775119 and 12175109), and the Natural Science Foundation of Tianjin (Grant  No: 19JCYBJC30800). {We are grateful to the referee for his constructive comments and suggestions.}

	\bibliographystyle{aasjournal}
	
	\bibliography{refer}

\begin{thebibliography}{}
\expandafter\ifx\csname natexlab\endcsname\relax\def\natexlab#1{#1}\fi
\providecommand{\url}[1]{\href{#1}{#1}}
\providecommand{\dodoi}[1]{doi:~\href{http://doi.org/#1}{\nolinkurl{#1}}}
\providecommand{\doeprint}[1]{\href{http://ascl.net/#1}{\nolinkurl{http://ascl.net/#1}}}
\providecommand{\doarXiv}[1]{\href{https://arxiv.org/abs/#1}{\nolinkurl{https://arxiv.org/abs/#1}}}

\bibitem[{Abadi {et~al.}(2016)Abadi, Barham, Chen, Chen, Davis, Dean, Devin,
  Ghemawat, Irving, Isard, {et~al.}}]{abadi16}
Abadi, M., Barham, P., Chen, J., {et~al.} 2016, arXiv:1605.08695

\bibitem[{Abbott {et~al.}(2017)Abbott, Abbott, Abbott, Acernese, Ackley, Adams,
  Adams, Addesso, Adhikari, Adya, {et~al.}}]{abbott17}
Abbott, B.~P., Abbott, R., Abbott, T., {et~al.} 2017, \prl, 119, 161101

\bibitem[{Abbott {et~al.}(2018)Abbott, Abbott, Abbott, Acernese, Ackley, Adams,
  Adams, Addesso, Adhikari, Adya, {et~al.}}]{abbott18}
---. 2018, \prl, 121, 161101

\bibitem[{Abbott {et~al.}(2019)Abbott, Abbott, Abbott, Acernese, Ackley, Adams,
  Adams, Addesso, Adhikari, Adya, {et~al.}}]{abbott19}
---. 2019, PhRvX, 9, 011001

\bibitem[{Abbott {et~al.}(2020)Abbott, Abbott, Abraham, Acernese, Ackley,
  Adams, Adhikari, Adya, Affeldt, Agathos, {et~al.}}]{abbott20b}
Abbott, R., Abbott, T., Abraham, S., {et~al.} 2020, \apjl, 896, L44

\bibitem[{Akmal {et~al.}(1998)Akmal, Pandharipande, \& Ravenhall}]{akmal98}
Akmal, A., Pandharipande, V.~R., \& Ravenhall, D. G.~a. 1998, \prc, 58, 1804

\bibitem[{Alvarez-Castillo {et~al.}(2016)Alvarez-Castillo, Ayriyan, Benic,
  Blaschke, Grigorian, \& Typel}]{alvarez16}
Alvarez-Castillo, D., Ayriyan, A., Benic, S., {et~al.} 2016, EPJA, 52, 1

\bibitem[{Antoniadis {et~al.}(2013)Antoniadis, Freire, Wex, Tauris, Lynch,
  Van~Kerkwijk, Kramer, Bassa, Dhillon, Driebe, {et~al.}}]{antoniadis13}
Antoniadis, J., Freire, P.~C., Wex, N., {et~al.} 2013, Sci., 340, 1233232

\bibitem[{Arzoumanian {et~al.}(2018)Arzoumanian, Brazier, Burke-Spolaor,
  Chamberlin, Chatterjee, Christy, Cordes, Cornish, Crawford, Cromartie,
  {et~al.}}]{arzoumanian18}
Arzoumanian, Z., Brazier, A., Burke-Spolaor, S., {et~al.} 2018, \apjs, 235, 37

\bibitem[{Bao {et~al.}(2014)Bao, Hu, Zhang, \& Shen}]{bao14}
Bao, S., Hu, J., Zhang, Z., \& Shen, H. 2014, \prc, 90, 045802

\bibitem[{Bao \& Shen(2015)}]{bao15}
Bao, S., \& Shen, H. 2015, \prc, 91, 015807

\bibitem[{Baym {et~al.}(2018)Baym, Hatsuda, Kojo, Powell, Song, \&
  Takatsuka}]{baym18}
Baym, G., Hatsuda, T., Kojo, T., {et~al.} 2018, RePP, 81, 056902

\bibitem[{Bender {et~al.}(2003)Bender, Heenen, \& Reinhard}]{bender03}
Bender, M., Heenen, P.-H., \& Reinhard, P.-G. 2003, RvModPh, 75, 121

\bibitem[{Chen {et~al.}(2013)Chen, Burgio, Schulze, \& Yasutake}]{chen13}
Chen, H., Burgio, G., Schulze, H.-J., \& Yasutake, N. 2013, \aap, 551, A13

\bibitem[{Chollet {et~al.}(2015)}]{chollet15}
Chollet, F., {et~al.} 2015, URL: https://keras. io/k, 7, T1

\bibitem[{Cromartie {et~al.}(2020)Cromartie, Fonseca, Ransom, Demorest,
  Arzoumanian, Blumer, Brook, DeCesar, Dolch, Ellis, {et~al.}}]{cromartie20}
Cromartie, H.~T., Fonseca, E., Ransom, S.~M., {et~al.} 2020, NatAs, 4, 72

\bibitem[{Cybenko(1989)}]{cybenko89}
Cybenko, G. 1989, MCSS, 2, 303

\bibitem[{Demorest {et~al.}(2010)Demorest, Pennucci, Ransom, Roberts, \&
  Hessels}]{demorest10}
Demorest, P.~B., Pennucci, T., Ransom, S., Roberts, M., \& Hessels, J. 2010,
  \nat, 467, 1081

\bibitem[{Dutra {et~al.}(2012)Dutra, Louren{\c{c}}o, Martins, Delfino, Stone,
  \& Stevenson}]{dutra12}
Dutra, M., Louren{\c{c}}o, O., Martins, J. S.~S., {et~al.} 2012, \prc, 85,
  035201

\bibitem[{Dutra {et~al.}(2014)Dutra, Louren{\c{c}}o, Avancini, Carlson,
  Delfino, Menezes, Provid{\^e}ncia, Typel, \& Stone}]{dutra14}
Dutra, M., Louren{\c{c}}o, O., Avancini, S., {et~al.} 2014, \prc, 90, 055203

\bibitem[{Essick {et~al.}(2020{\natexlab{a}})Essick, Landry, \&
  Holz}]{essick20a}
Essick, R., Landry, P., \& Holz, D.~E. 2020{\natexlab{a}}, \prd, 101, 063007

\bibitem[{Essick {et~al.}(2020{\natexlab{b}})Essick, Tews, Landry, Reddy, \&
  Holz}]{essick20b}
Essick, R., Tews, I., Landry, P., Reddy, S., \& Holz, D.~E. 2020{\natexlab{b}},
  \prd, 102, 055803

\bibitem[{Farrell {et~al.}(2022)Farrell, Baldi, Ott, Ghosh, Steiner, Kavitkar,
  Lindblom, Whiteson, \& Weber}]{farrell22}
Farrell, D., Baldi, P., Ott, J., {et~al.} 2022, arXiv:2209.02817

\bibitem[{Fattoyev {et~al.}(2020)Fattoyev, Horowitz, Piekarewicz, \&
  Reed}]{bigapple}
Fattoyev, F., Horowitz, C., Piekarewicz, J., \& Reed, B. 2020, \prc, 102,
  065805

\bibitem[{Ferreira {et~al.}(2022)Ferreira, Carvalho, \&
  Provid{\^e}ncia}]{ferreira22}
Ferreira, M., Carvalho, V., \& Provid{\^e}ncia, C. 2022, arXiv:2209.09085

\bibitem[{Ferreira \& Provid{\^e}ncia(2021)}]{ferreira21}
Ferreira, M., \& Provid{\^e}ncia, C. 2021, JCAP, 2021, 011

\bibitem[{Fonseca {et~al.}(2016)Fonseca, Pennucci, Ellis, Stairs, Nice, Ransom,
  Demorest, Arzoumanian, Crowter, Dolch, {et~al.}}]{fonseca16}
Fonseca, E., Pennucci, T.~T., Ellis, J.~A., {et~al.} 2016, \apj, 832, 167

\bibitem[{Fujimoto {et~al.}(2018)Fujimoto, Fukushima, \& Murase}]{fujimoto18}
Fujimoto, Y., Fukushima, K., \& Murase, K. 2018, \prd, 98, 023019

\bibitem[{Fujimoto {et~al.}(2020)Fujimoto, Fukushima, \& Murase}]{fujimoto20}
---. 2020, \prd, 101, 054016

\bibitem[{Fujimoto {et~al.}(2021)Fujimoto, Fukushima, \& Murase}]{fujimoto21}
---. 2021, JHEP, 2021, 1

\bibitem[{Glendenning(2001)}]{glendenning01}
Glendenning, N.~K. 2001, PhR, 342, 393

\bibitem[{Glorot \& Bengio(2010)}]{glorot10}
Glorot, X., \& Bengio, Y. 2010, JMLR Workshop and Conference Proceedings
  (Proceedings of the thirteenth international conference on artificial
  intelligence and statistics), 249--256

\bibitem[{Han {et~al.}(2021)Han, Jiang, Tang, \& Fan}]{han21}
Han, M.-Z., Jiang, J.-L., Tang, S.-P., \& Fan, Y.-Z. 2021, \apj, 919, 11

\bibitem[{Hornik(1991)}]{hornik91}
Hornik, K. 1991, Neural Netw., 4, 251

\bibitem[{Hu {et~al.}(2020)Hu, Bao, Zhang, Nakazato, Sumiyoshi, \& Shen}]{hu20}
Hu, J., Bao, S., Zhang, Y., {et~al.} 2020, PTEP, 2020, 043D01

\bibitem[{Huang {et~al.}(2020)Huang, Hu, Zhang, \& Shen}]{huang20}
Huang, K., Hu, J., Zhang, Y., \& Shen, H. 2020, \apj, 904, 39

\bibitem[{Huang {et~al.}(2022)Huang, Hu, Zhang, \& Shen}]{huang22}
---. 2022, \apj, 935, 88

\bibitem[{Ji {et~al.}(2019)Ji, Hu, Bao, \& Shen}]{ji19}
Ji, F., Hu, J., Bao, S., \& Shen, H. 2019, \prc, 100, 045801

\bibitem[{Ju {et~al.}(2021)Ju, Wu, Ji, Hu, \& Shen}]{ju21}
Ju, M., Wu, X., Ji, F., Hu, J., \& Shen, H. 2021, \prc, 103, 025809

\bibitem[{Kingma \& Ba(2014)}]{kingma14}
Kingma, D.~P., \& Ba, J. 2014, arXiv:1412.6980

\bibitem[{Lalazissis {et~al.}(1997)Lalazissis, K{\"o}nig, \& Ring}]{nl3}
Lalazissis, G., K{\"o}nig, J., \& Ring, P. 1997, \prc, 55, 540

\bibitem[{Lalazissis {et~al.}(2005)Lalazissis, Nik{\v{s}}i{\'c}, Vretenar, \&
  Ring}]{ddme2}
Lalazissis, G., Nik{\v{s}}i{\'c}, T., Vretenar, D., \& Ring, P. 2005, \prc, 71,
  024312

\bibitem[{Landry \& Essick(2019)}]{landry19}
Landry, P., \& Essick, R. 2019, \prd, 99, 084049

\bibitem[{Lattimer \& Prakash(2000)}]{lattimer00}
Lattimer, J.~M., \& Prakash, M. 2000, PhR, 333, 121

\bibitem[{Lattimer \& Prakash(2007)}]{lattimer07}
---. 2007, PhR, 442, 109

\bibitem[{Li {et~al.}(2008)Li, Chen, \& Ko}]{li08}
Li, B.-A., Chen, L.-W., \& Ko, C.~M. 2008, PhR, 464, 113

\bibitem[{Li {et~al.}(2019)Li, Krastev, Wen, \& Zhang}]{li19}
Li, B.-A., Krastev, P.~G., Wen, D.-H., \& Zhang, N.-B. 2019, EPJA, 55, 1

\bibitem[{Lindblom(1992)}]{lindblom92}
Lindblom, L. 1992, \apj, 398, 569

\bibitem[{Lindblom(2010)}]{lindblom10}
---. 2010, \prd, 82, 103011

\bibitem[{Long {et~al.}(2004)Long, Meng, Van~Giai, \& Zhou}]{pkdd}
Long, W., Meng, J., Van~Giai, N., \& Zhou, S.-G. 2004, \prc, 69, 034319

\bibitem[{Meng {et~al.}(2006)Meng, Toki, Zhou, Zhang, Long, \& Geng}]{meng06}
Meng, J., Toki, H., Zhou, S.-G., {et~al.} 2006, PPNuPh, 57, 470

\bibitem[{Miao {et~al.}(2021)Miao, Jiang, Li, \& Chen}]{miao21}
Miao, Z., Jiang, J.-L., Li, A., \& Chen, L.-W. 2021, \apjl, 917, L22

\bibitem[{Miller {et~al.}(2019)Miller, Lamb, Dittmann, Bogdanov, Arzoumanian,
  Gendreau, Guillot, Harding, Ho, Lattimer, {et~al.}}]{miller19}
Miller, M.~C., Lamb, F.~K., Dittmann, A.~J., {et~al.} 2019, \apjl, 887, L24

\bibitem[{Miller {et~al.}(2021)Miller, Lamb, Dittmann, Bogdanov, Arzoumanian,
  Gendreau, Guillot, Ho, Lattimer, Loewenstein, {et~al.}}]{miller21}
---. 2021, \apjl, 918, L28

\bibitem[{Murarka {et~al.}(2022)Murarka, Banerjee, Malik, \&
  Provid{\^e}ncia}]{murarka22}
Murarka, U., Banerjee, K., Malik, T., \& Provid{\^e}ncia, C. 2022, JCAP, 2022,
  045

\bibitem[{Nik{\v{s}}i{\'c} {et~al.}(2002)Nik{\v{s}}i{\'c}, Vretenar, Finelli,
  \& Ring}]{ddme1}
Nik{\v{s}}i{\'c}, T., Vretenar, D., Finelli, P., \& Ring, P. 2002, \prc, 66,
  024306

\bibitem[{Nik{\v{s}}i{\'c} {et~al.}(2011)Nik{\v{s}}i{\'c}, Vretenar, \&
  Ring}]{niksic11}
Nik{\v{s}}i{\'c}, T., Vretenar, D., \& Ring, P. 2011, PPNuPh, 66, 519

\bibitem[{Oertel {et~al.}(2017)Oertel, Hempel, Kl{\"a}hn, \& Typel}]{oertel17}
Oertel, M., Hempel, M., Kl{\"a}hn, T., \& Typel, S. 2017, RvModPh, 89, 015007

\bibitem[{Oppenheimer \& Volkoff(1939)}]{oppenheimer39}
Oppenheimer, J.~R., \& Volkoff, G.~M. 1939, PhRe, 55, 374

\bibitem[{Orsaria {et~al.}(2014)Orsaria, Rodrigues, Weber, \&
  Contrera}]{orsaria14}
Orsaria, M., Rodrigues, H., Weber, F., \& Contrera, G. 2014, \prc, 89, 015806

\bibitem[{{\"O}zel {et~al.}(2010){\"O}zel, Baym, \& G{\"u}ver}]{ozel10}
{\"O}zel, F., Baym, G., \& G{\"u}ver, T. 2010, \prd, 82, 101301

\bibitem[{Raithel {et~al.}(2017)Raithel, {\"O}zel, \& Psaltis}]{raithel17}
Raithel, C.~A., {\"O}zel, F., \& Psaltis, D. 2017, \apj, 844, 156

\bibitem[{Read {et~al.}(2009)Read, Lackey, Owen, \& Friedman}]{read09}
Read, J.~S., Lackey, B.~D., Owen, B.~J., \& Friedman, J.~L. 2009, Phys. Rev. D,
  79, 124032, \dodoi{10.1103/PhysRevD.79.124032}

\bibitem[{Riley {et~al.}(2019)Riley, Watts, Bogdanov, Ray, Ludlam, Guillot,
  Arzoumanian, Baker, Bilous, Chakrabarty, {et~al.}}]{riley19}
Riley, T.~E., Watts, A.~L., Bogdanov, S., {et~al.} 2019, \apjl, 887, L21

\bibitem[{Riley {et~al.}(2021)Riley, Watts, Ray, Bogdanov, Guillot, Morsink,
  Bilous, Arzoumanian, Choudhury, Deneva, {et~al.}}]{riley21}
Riley, T.~E., Watts, A.~L., Ray, P.~S., {et~al.} 2021, \apjl, 918, L27

\bibitem[{Ring(1996)}]{ring96}
Ring, P. 1996, PPNuPh, 37, 193

\bibitem[{Sammarruca(2010)}]{sammarruca10}
Sammarruca, F. 2010, IJModPhE, 19, 1259

\bibitem[{Sammarruca {et~al.}(2012)Sammarruca, Chen, Coraggio, Itaco, \&
  Machleidt}]{sammarruca12}
Sammarruca, F., Chen, B., Coraggio, L., Itaco, N., \& Machleidt, R. 2012, \prc,
  86, 054317

\bibitem[{Steiner {et~al.}(2010)Steiner, Lattimer, \& Brown}]{steiner10}
Steiner, A.~W., Lattimer, J.~M., \& Brown, E.~F. 2010, \apj, 722, 33

\bibitem[{Stone \& Reinhard(2007)}]{stone07}
Stone, J.~R., \& Reinhard, P.-G. 2007, PPNuPh, 58, 587

\bibitem[{Sun(2016)}]{sun16}
Sun, B. 2016, Sci. Sin. Phys., Mech. Astron., 46, 012018

\bibitem[{Taninah {et~al.}(2020)Taninah, Agbemava, Afanasjev, \& Ring}]{ddmex}
Taninah, A., Agbemava, S., Afanasjev, A., \& Ring, P. 2020, PhLB, 800, 135065

\bibitem[{Tolman(1939)}]{tolman39}
Tolman, R.~C. 1939, PhRe, 55, 364

\bibitem[{Typel {et~al.}(2010)Typel, R{\"o}pke, Kl{\"a}hn, Blaschke, \&
  Wolter}]{dd2}
Typel, S., R{\"o}pke, G., Kl{\"a}hn, T., Blaschke, D., \& Wolter, H. 2010,
  \prc, 81, 015803

\bibitem[{Van~Dalen {et~al.}(2004)Van~Dalen, Fuchs, \& Faessler}]{dalen04}
Van~Dalen, E., Fuchs, C., \& Faessler, A. 2004, NuPhA, 744, 227

\bibitem[{Wang {et~al.}(2020)Wang, Hu, Zhang, \& Shen}]{wang20}
Wang, C., Hu, J., Zhang, Y., \& Shen, H. 2020, \apj, 897, 96

\bibitem[{Weber(2005)}]{weber05}
Weber, F. 2005, PPNuPh, 54, 193

\bibitem[{Wei {et~al.}(2020)Wei, Zhao, Wang, Geng, Sun, Niu, \& Long}]{ddlz1}
Wei, B., Zhao, Q., Wang, Z.-H., {et~al.} 2020, CPhC, 44, 074107

\bibitem[{Wei {et~al.}(2019)Wei, Figura, Burgio, Chen, \& Schulze}]{wei19}
Wei, J.~B., Figura, A., Burgio, G.~F., Chen, H., \& Schulze, H. 2019, JPhG, 46,
  034001

\bibitem[{Williams \& Rasmussen(2006)}]{williams06}
Williams, C.~K., \& Rasmussen, C.~E. 2006 (MIT press Cambridge, MA)

\bibitem[{Wu \& Shen(2017)}]{wu17}
Wu, X., \& Shen, H. 2017, \prc, 96, 025802

\bibitem[{Xu {et~al.}(2010)Xu, Chen, Ko, \& Li}]{xu10}
Xu, J., Chen, L.-W., Ko, C.~M., \& Li, B.-A. 2010, \prc, 81, 055803

\bibitem[{Yang \& Shen(2008)}]{yang08}
Yang, F., \& Shen, H. 2008, \prc, 77, 025801

\end{thebibliography}
	\listofchanges 
\end{document}